\documentclass[epj]{svjour}

\usepackage{graphicx}
\usepackage{fancyhdr}
\def\makeheadbox{{
 }}
\usepackage{amsmath,amssymb,amsfonts,color,cite,color}
\usepackage{colordvi,psfrag} 
\input paperdef
\def\mytitle{My title} 
\def\myauthors{My name}  
\def\mytype{My type of session}
\def\mysession{My session}

\def\mytitle{Higgs boson decays in the Complex MSSM} 
\def\myauthors{K Williams, G Weiglein}    
\def\mytype{Contributed Talk}    
\def\mysession{Colliders - Higgs Phenomenology}

\begin{document}
\title{Higgs boson decays in the Complex MSSM}
\author{Karina Williams\inst{1}
\thanks{\emph{Email: k.e.williams@durham.ac.uk}}%
 \and
 Georg Weiglein\inst{1} 
\thanks{\emph{Email: georg.weiglein@durham.ac.uk}}%
}                     
%
%
\institute{IPPP, University of Durham, Durham DH1~3LE, UK}
%
\date{}
\abstract{
The analysis of the Higgs search results at LEP showed that a part 
of the MSSM parameter space with non-zero complex phases could not 
be excluded, where the lightest neutral Higgs boson, $h_1$, has a mass
of only about 45~GeV and the second lightest neutral Higgs boson, $h_2$, has a 
sizable branching fraction into a pair of $h_1$ states. Full one-loop
results for the Higgs cascade decay $h_2 \to h_1h_1$ are presented and combined 
with two-loop Higgs propagator corrections taken from the program \fh.
Using the improved theoretical prediction to analyse the limits on topological cross
sections obtained at LEP, the existence of an unexcluded region at low Higgs mass is 
confirmed. The effect of the genuine vertex corrections on the size and shape of this region 
is discussed.
} 
\maketitle
%

\section{Introduction}
\label{intro}

In its general form, the MSSM contains complex phases which drive $\cp$-violation.  
The results of the Higgs-boson searches at LEP were analysed in a series of MSSM benchmark 
scenarios~\cite{benchmark,cpx}, including one explicitly chosen to investigate the effects 
of these $\cp$-violating phases, called the CPX scenario~\cite{cpx}.
This scenario produced an unexcluded region for light $h_1$ and relatively small values of $\tb$ (the ratio of the vacuum expectation values of the two Higgs doublets), so that no firm lower bound on the mass of the lightest Higgs boson of the MSSM could be set~\cite{LEPHiggs}. Throughout much of this region of CPX parameter space, the process $h_2 \to h_1h_1$ dominates the $h_2$ decay width. 
Therefore, in order to reliably determine which parameter regions of the MSSM are unexcluded by the Higgs searches so far and which regions will be accessible by Higgs searches in the future, precise predictions for this process are indispensable. 

However, the genuine vertex corrections (as opposed to propagator corrections) to the process $h_2 \to h_1h_1$ in
the MSSM with complex parameters have so far been unavailable within the Feynman-diagrammatic (FD) approach. These vertex corrections are expected to be 
sizable as they contain terms depending on the fourth power of the top mass.
We have obtained complete one-loop results within the FD approach
for the decays $h_a \to h_b h_c$, including all genuine one-loop vertex corrections, in the MSSM with complex phases (cMSSM)~\cite{hdeccpv}. 
We have furthermore calculated complete one-loop results for the decays of neutral Higgs bosons into SM fermions in the cMSSM~\cite{hdeccpv}. 
These new one-loop results are combined with existing higher-order corrections in the FD approach and will be included in the public code 
\fh~\cite{fhrandproc,feynhiggs,mhcMSSMlong,mhcpv2l}. As an application of our improved result, we analyse the coverage of the LEP exclusion limits on the 
topological cross sections for the various Higgs-boson production and decay 
cross sections within the $\MHe$--$\tb$ parameter plane. 

\section{Calculation of Higgs-boson cascade decays
  and decays into SM fermions}
\label{sec:1}

We calculated the full 1PI (one-particle irreducible) one-loop vertex
contributions to the processes $h_a \rightarrow h_b h_c$, taking into
account all sectors of the MSSM and the complete dependence on the
complex phases of the supersymmetric parameters. These calculations (which make use of the programs \fa~\cite{feynarts} and \fc~\cite{formcalc}) have
been described in more detail in \citere{hdeccpv}.

We have derived also
complete one-loop results for the processes $h_a \to f \bar{f}$ 
(including SM-type QED and, where appropriate, QCD corrections) for
arbitrary values of all complex phases of the supersymmetric parameters, as these decays are important over large parts of the cMSSM parameter space.
The partial decay widths for the other Higgs-boson decay modes have been taken
from the program \fh.

In our numerical analysis below, we will compare our full result with the contribution from just the $t,\tilde{t}$ sector in the approximation where the gauge couplings are neglected and the diagrams are evaluated at zero external momenta, which we call the `Yukawa Approximation'.

For the renormalisation of the Higgs fields it is convenient to use a \drbar\
scheme as in \citere{mhcMSSMlong}. Therefore, it is necessary to include finite wave-function normalisation factors (Z-factors) to ensure that S-matrix
elements involving external Higgs fields have the correct on-shell properties. The form of these Z-factors is given in \citere{hdeccpv}. 
In the calculation of the Z-factors, we incorporate higher-order contributions obtained from the neutral Higgs-boson self-energies of the program \fh. 
In the analysis below, we will restrict to those higher-order contributions for which the phase dependence at the two-loop is explicitly 
known~\cite{mhcpv2l}, i.e.\
we do not include contributions that have been extrapolated from results obtained for the real MSSM.

\section{Implementation of exclusion bounds from the LEP Higgs searches}

\label{sec:2}

In our numerical analysis below we analyse the impact of our new result 
on the LEP exclusion bounds in the cMSSM. This is done by comparing
the cMSSM predictions with the topological cross section limits given in
\citeres{LEPHiggs,philip}. This procedure is described in more detail in \citere{hdeccpv} and \citere{higgsbounds}. It involves checking which channel is predicted to have the highest statistical sensitivity and then comparing the theoretical prediction in that channel only with the topological cross section limit determined at LEP. This ensures a correct statistical interpretation of the overall exclusion limit at the 95\%
C.L.. However, it should be noted that, since only one channel can be used at a time, this method leads to poorer coverage in areas where two or more channels have a similar statistical sensitivity than, for example, the dedicated analyses carried out in \citere{LEPHiggs}.

\section{Numerical results}
\label{sec:3}

In our numerical analysis we use the parameter values of the CPX
benchmark scenario~\cite{cpx}, adapted to the latest experimental
central value of the top-quark mass~\cite{mt1709} and using an on-shell
value for the absolute value of the trilinear couplings 
$\At$ and $\Ab$ 
that is somewhat shifted compared to the 
\drbar\ value specified in \citere{cpx}. Specifically, if not
indicated differently, we use the parameters
\begin{align}
& \msusy = 500 \gev, \; |\At| = |\Ab| = 900 \gev, \non \\
&  \mu = 2000 \gev, \; \mgl \equiv |M_3| = 1000 \gev , \non \\
& \mt = 170.9 \gev, \; M_2 = 200 \gev .
\label{eq:cpx1}
\end{align}
The lowest-order Higgs sector parameters 
$\tb$ and $\MHp$ are varied, with $\MHp<1000$. The complex phases of the trilinear couplings $\At$, $\Ab$
and the gluino mass parameter $M_3$ are set to
\begin{equation}
\phiat = \frac{\pi}{2}, \quad  \phiab = \frac{\pi}{2}, 
\quad \phigl = \frac{\pi}{2} .
\label{eq:cpx2}
\end{equation}
In \refeq{eq:cpx1}
$\msusy$ denotes the diagonal soft SUSY-breaking parameters in the
sfermion mass matrices, which are chosen to be equal to each other.

\subsection{Results for the $h_2\rightarrow h_1h_1$ decay width}

\begin{figure}
\begin{center}
\includegraphics[height=0.6\linewidth,angle=0]{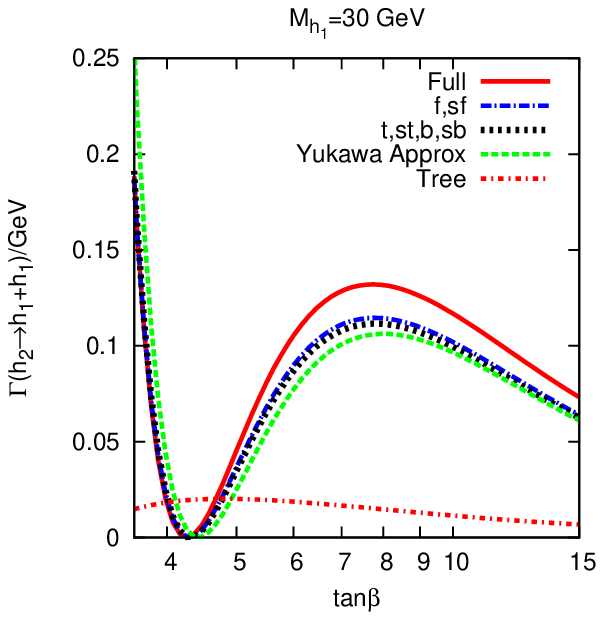}
\end{center}
\caption{ $\Ga(h_2 \to h_1h_1)$ as function of $\tb$ 
($\MHp$ is adjusted to ensure $\MHe = 30 \gev$), for various approximations.
\label{linegraphs}
}
\end{figure}

\begin{figure}
\begin{center}
\includegraphics[height=0.70\linewidth,angle=0]{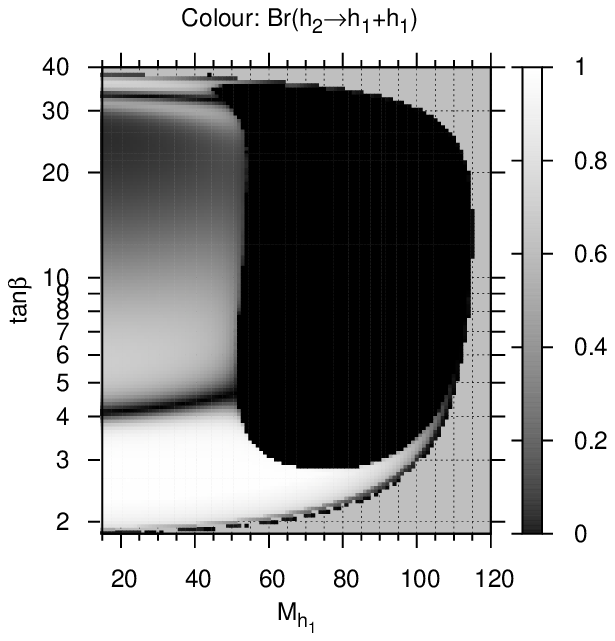}
\end{center}
\vspace{-2em}
\caption{The branching ratio $\br{(h_2\to h_1h_1)}$ in the $\MHe$--$\tb$
plane of the CPX scenario.
\label{MHpTBplots1}
}
\end{figure}

\reffi{linegraphs} shows the relative effect of different contributions to 
the $h_2\rightarrow h_1h_1$ decay width in an area of the cMSSM parameter
space that is particularly relevant to the unexcluded region in the LEP Higgs searches. The decay width is plotted against $\tb$ whilst 
adjusting $\MHp$ such that $\MHe = 30 \gev$. All the results plotted include the higher-order corrected wave-function
normalisation factors as described in \refse{sec:2}. The results differ only in the genuine contributions to the $h_2h_1h_1$ vertex. One can see that the full result (denoted as
`Full') differs drastically from the case where only wave-function
normalisation factors but no genuine one-loop vertex contributions are
taken into account (`Tree'). The Yukawa
approximation agrees much better with the full result and explains the qualitative behaviour of the full result --- 
in particular, the region where $\Ga(h_2 \to h_1h_1) \approx 0$ for $\tb \approx 4.3$
is due to the fact that the Yukawa vertex corrections to the matrix
element change sign when $\tb$ is varied. Using the full contribution from the $t,\tilde{t},b,\tilde{b}$ sector (`t, st, b, sb') or 
using all three generations of fermions and sfermions (`f, sf') yields a prediction that deviates from the full result by up to 15\%.

\begin{figure}
\begin{center}
\includegraphics[height=0.7\linewidth,angle=0]{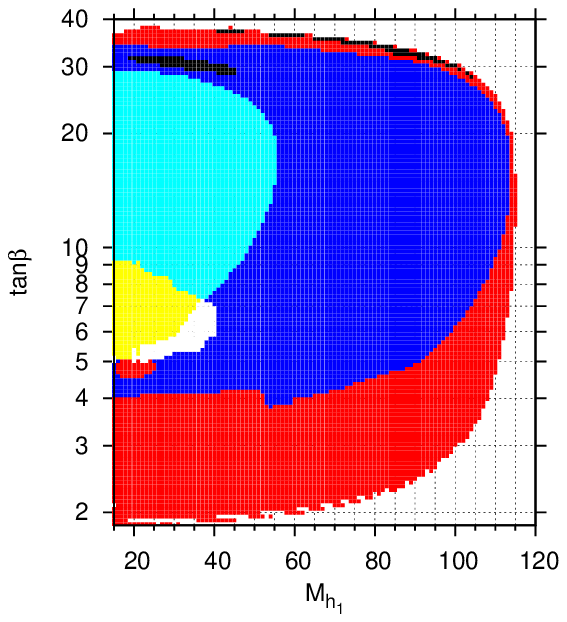}
\end{center}
\caption{The channels predicted to have the highest statistical sensitivity for setting an
exclusion limit using the results from the LEP Higgs searches in the CPX scenario. 
The colour codings are: red $=$ $(h_1 Z)\to(b \bar b Z)$,
blue $=$ $(h_2  Z)\to(b \bar b Z)$,
white $=$ $(h_2  Z)\to(h_1 h_1 Z)\to(b \bar b b \bar b Z)$,
cyan $=$ $(h_2 h_1)\to(b \bar b b \bar b)$,
yellow $=$ $(h_2 h_1)\to(h_1 h_1 h_1)\to(b \bar b b \bar b b \bar b)$,
black $=$ other channels.
\label{MHpTBplots2}
}
\end{figure}

\begin{figure}
\begin{center}
\includegraphics[height=0.7\linewidth,angle=0]{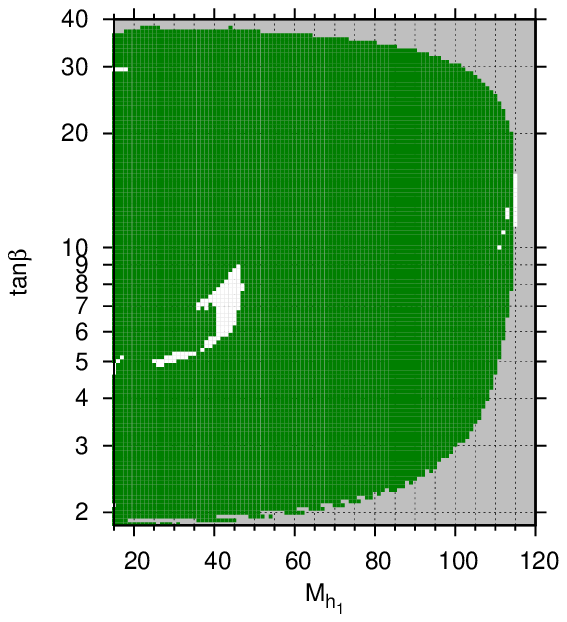}
\end{center}
\caption{The parameter region in the CPX scenario excluded 
at the 95\% C.L.\
by the topological cross section limits obtained at LEP~\cite{LEPHiggs}.
The colour codings are:
green = LEP excluded, white = LEP allowed, grey = theoretically
inaccessible.
\label{MHpTBplots3}
}
\end{figure}

In \reffis{MHpTBplots1}, \ref{MHpTBplots2} and \ref{MHpTBplots3} the $\MHe$--$\tb$ parameter 
space of the CPX scenario is analysed. \reffi{MHpTBplots1} shows the
branching ratio of the Higgs
cascade decay, $\br{(h_2\to h_1h_1)}$. Over a large
part of the parameter space where the decay $h_2\to h_1h_1$ is
kinematically possible it is actually the dominant decay channel. The
branching ratio is particularly large for low and moderate values of
$\tb$. In the region where $\tb \approx 4$--5 the Yukawa contribution to
the matrix element for the $h_2\to h_1h_1$ decay changes sign
and causes a sharp drop in the $h_2\to h_1h_1$ branching ratio,
as was already observed in \reffi{linegraphs}. A similar behaviour 
occurs also in the region of $\tb \approx 35$. 

\reffi{MHpTBplots2} indicates which channel has the highest
statistical sensitivity and therefore which channel will be used to set an exclusion limit in different regions of the parameter space. As explained in
\refse{sec:3}, this information is needed for
a statistical interpretation of the topological cross section limits
obtained at LEP. 
One can see in the figure that the
channels $e^+e^- \to h_2  Z\to  h_1 h_1 Z \to b \bar b b \bar b Z$
and $e^+e^- \to h_2 h_1 \to h_1 h_1 h_1 \to b \bar b b \bar b b \bar b$
have the highest search sensitivity in a region with small $\MHe$ and
small and moderate values of $\tb$ (the region at $\tb \approx 4$ where
the channel $e^+e^- \to(h_2  Z)\to(b \bar b Z)$ has the highest search
sensitivity corresponds to the drop in the
$h_2\to h_1h_1$ branching ratio observed in \reffi{MHpTBplots1}).
For small $\MHe$ and somewhat higher $\tb$ the channel
$e^+e^- \to (h_2 h_1)\to(b \bar b b \bar b)$ has the highest search
sensitivity. It should be noted that all channels involving the decay of
the $h_2$ boson in the region of small $\MHe$ are strongly influenced by
the $\Ga(h_2 \to h_1h_1)$ decay width, either directly in the case of
the channels involving the Higgs cascade decay, or indirectly through
the branching ratio of the $h_2$.
The parameter region where $\Ga(h_2 \to h_1h_1)$
is important coincides with the
region of the CPX scenario that could not be excluded at the 95\% C.L.\
in the analysis of the four LEP collaborations~\cite{LEPHiggs}.

In \reffi{MHpTBplots3} we have compared our new theoretical
predictions 
with the topological cross section limits obtained at LEP
for the channels in \reffi{MHpTBplots2}.
We find an unexcluded region at $\MHe \approx 45 \gev$ and moderate
$\tb$ where channels involving the decay $h_2 \to h_1h_1$ play an
important role. Thus, our analysis, based on the most
up-to-date theory prediction for the $h_2 \to h_1h_1$ channel, confirms
the `hole' in the LEP coverage observed in \citere{LEPHiggs}
(see in particular Fig.~19 of \citere{LEPHiggs}).

It should be noted, on the other hand, that the results shown in \reffi{MHpTBplots3} differ in several respects from the results presented in \citere{LEPHiggs}. As discussed above, our analysis has less
statistical sensitivity near borders between areas where different search topologies 
are predicted to have the highest exclusion power than the benchmark
studies of \citere{LEPHiggs}. A further difference (besides the improved
theoretical prediction)
is the input value of the top-quark mass. While
we are using the latest experimental central value of 
$\mt = 170.9$~\cite{mt1709}, most of the analysis of \citere{LEPHiggs} was
done for $\mt = 174.3$. We have explicitly checked that (as expected)
the unexcluded region is significantly increased if we use 
$\mt = 174.3$ instead. It should also be noted that the shape and size of the unexcluded region depends sensitively on the value of $A_t$ used.

\begin{figure*}
\begin{center}
\begin{tabular}{ccc}
\includegraphics[height=0.25\linewidth,angle=0]{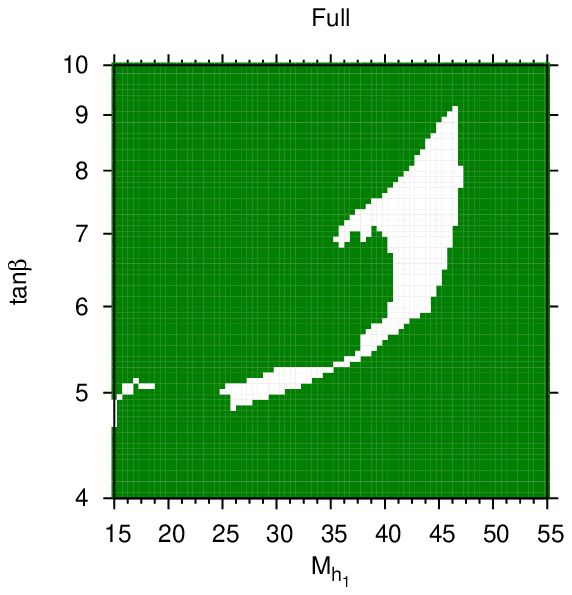}&
\includegraphics[height=0.25\linewidth,angle=0]{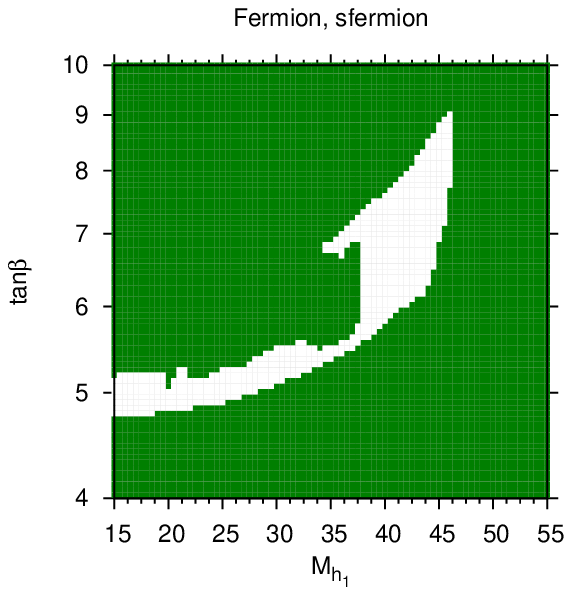}&
\includegraphics[height=0.25\linewidth,angle=0]{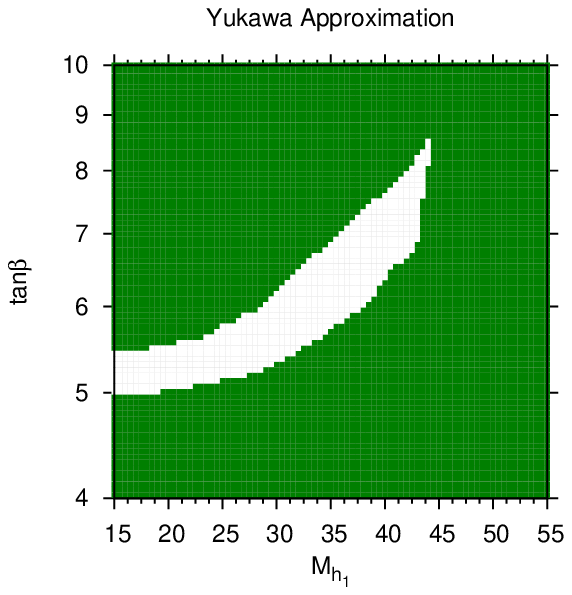}\\
(a)&(b)&(c)
\end{tabular}
\end{center}
\caption{A closer look at the LEP coverage for 
$\MHe \approx 45 \gev$ and moderate $\tb$. Plot (a) shows the full
result (detailed view of plot (b) of \reffi{MHpTBplots2}). Plot (b)
shows the result for the case where only contributions from SM fermions
and their superpartners are taken into account in the genuine vertex 
corrections. Plot (c) shows the result where the Yukawa Approximation
has been used for the genuine vertex corrections.
The colour codings are:
green = LEP excluded, white = LEP allowed.
\label{approxLEP}
}
\end{figure*}

In \reffi{approxLEP} we focus on the parameter region at 
$\MHe \approx 45 \gev$ and moderate $\tb$. While plot (a) shows the full result (i.e., it is a 
detailed view of \reffi{MHpTBplots3}), in plot (b) the
genuine $h_2\to h_1h_1$ vertex corrections are approximated by the contributions from
fermions and sfermions only, and in plot (c) the Yukawa Approximation
has been used for the genuine $h_2\to h_1h_1$ vertex corrections. In all three plots the
wave function normalisation factors and all other decay widths are the
same. While all three graphs show the unexcluded region to be in a similar position in 
parameter space, the shape of this unexcluded region changes quite significantly.

\section{Conclusions}
\label{sec:4}

We have shown that our new result for the decay width $h_2\to h_1h_1$, which includes the complete one-loop vertex corrections, drastically improves on
results obtained using propagator corrections alone. We have combined this with new results for neutral Higgs decays into two fermions in order to
re-analyse the CPX parameter space using the limits on topological cross sections obtained from the LEP Higgs searches. We find that over a large part of
the parameter space where the decay $h_2\to h_1h_1$ is kinematically possible, it is the dominant decay channel. We confirm that the parameter space at
low Higgs mass can not be completely excluded by the 
LEP Higgs searches. We furthermore show that the size and shape of the unexcluded region is significantly modified if approximations to the full result
are used, emphasizing the relevance of the full one-loop result for the $h_2\to h_1h_1$ decay width in the complex MSSM.

\subsection*{Acknowledgements}

We are grateful to P.~Bechtle, O.~Brein, T.~Hahn, S.~Heinemeyer, W.~Hollik,
S.~Palmer, H.~Rzehak and D.~St\"ockinger for many helpful discussions.
Work supported in part by the European Community's Marie-Curie Research
Training Network under contract MRTN-CT-2006-035505
`Tools and Precision Calculations for Physics Discoveries at Colliders'.

%
%

\end{document}